# The Longevity of Innovation

Yiling Lin[1], Zak Risha[1], Erin Leahey[2]*, Lingfei Wu[1]*

**Affiliations**

[1] School of Computing and Information, University of Pittsburgh, Pittsburgh, PA, USA

[2] School of Sociology, University of Arizona, 1145 E. South Campus Drive, Tucson, AZ, USA

*Corresponding authors. Erin Leahey (leahey@arizona.edu) and Lingfei Wu (liw105@pitt.edu)

**Abstract (Maximum: 250 words)**

Modern science is organized around specialization in training and teamwork. Scientists develop deep expertise within a field and combine complementary knowledge through collaboration to solve complex problems. Yet whether specialization is the most effective path to sustained innovation remains unclear. Here we introduce a quantitative framework that distinguishes generalists from specialists based on scaling patterns of disciplinary mobility while remaining independent of career age and productivity. Applying this framework to 49 million publications produced by 3 million scientists between 1900 and 2020, we examine how research style relates to innovation, learning, collaboration, and productivity. We find that scientists who move across fields are more likely to sustain innovative contributions throughout their careers, whereas those who remain within narrow fields exhibit the age-related decline in innovation. Generalists are less anchored to the literature of their training. They are more likely to pursue research independently, and, when they collaborate, they preferentially partner with other generalists. Teams with a greater share of generalists produce more innovative research, even after accounting for differences in knowledge diversity. Despite these advantages, generalists publish fewer papers on average and have become less common over time. These findings reveal a tension between the longevity of scientific careers and the longevity of scientific innovation.

## Introduction

Innovation is central to scientific progress. Yet how innovation changes over the scientific life course remains debated. Some studies suggest that innovation declines with age (1–3), whereas others emphasize the benefits of accumulated expertise and experience (4–6). Despite their differences, these perspectives share a common assumption: that scientists follow similar innovation trajectories throughout their careers. Yet an important question remains: Why do some scientists continue to innovate throughout their careers while others do not?

Existing explanations largely emphasize the accumulation of knowledge (4–6). Deep expertise is often viewed as the foundation of innovation because new ideas build upon existing ones (7–9). Yet this view presents a puzzle. If knowledge accumulation alone drives innovation, scientists should become increasingly innovative as they deepen their expertise within familiar domains. Instead, innovation often declines with age (1–3). This suggests that sustained innovation may require more than the accumulation of knowledge. It may also depend on the ability to renew the intellectual foundations of one's work.

This possibility resonates with a long tradition of scholarship emphasizing that innovation requires the displacement of existing ideas. Schumpeter described economic growth as a process of "creative destruction" in which new technologies replace old ones (10). Building on this foundation, subsequent theories of innovation-driven growth likewise emphasized progress through successive waves of technological change (11). Historians and philosophers of science reached similar conclusions. Popper argued that scientific progress depends on the rejection of existing theories (12), whereas Kuhn emphasized paradigm shifts that transform prevailing frameworks of understanding (13). Across these perspectives, innovation is sustained not only through the creation of new ideas but also through the revision, displacement, or abandonment of old ones.

Yet these theories typically locate renewal at the level of systems rather than individuals. Technologies replace technologies, firms replace firms, paradigms replace paradigms, and generations replace generations. When it comes to individuals, however, the prevailing view is far more pessimistic. Christensen argued that disruptive innovation often requires new organizations rather than adaptation by existing ones (14). Planck famously suggested that science advances "one funeral at a time," implying that intellectual renewal occurs through generational replacement rather than individual change (15). Recent studies similarly document tensions between established and emerging ideas across scientific generations (2, 3). Together, these perspectives suggest that systems may renew themselves even when individuals remain attached to the intellectual foundations formed early in their careers (3). Whether individuals themselves can repeatedly renew those foundations, and thereby sustain innovation over time, remains largely unknown.

One possible mechanism for individual intellectual renewal is knowledge mobility. Scientists differ in how broadly they explore knowledge throughout their careers (16, 17). Some repeatedly enter new intellectual communities and pursue unfamiliar problems, whereas others remain concentrated within established domains. Existing explanations emphasize the benefits of knowledge accumulation and often assume that scientists who move across fields are more innovative because they bring more diverse experiences to their work (18–20). Yet mobility may offer another advantage. By repeatedly entering unfamiliar domains, scientists may become less anchored to the literature of their training. Each move places them in the position of an intellectual newcomer, exposing them to new ideas, collaborators, and ways of thinking while weakening attachment to established intellectual commitments. If so, knowledge mobility may help explain why some scientists sustain innovation throughout their careers while others do not.

To investigate this possibility, we introduce a quantitative framework that distinguishes generalists from specialists using scaling patterns of field mobility independent of career age and productivity. The resulting scale-free parameter captures persistent differences in how scientists explore knowledge throughout their careers. Applying this framework to 49 million publications produced by 3 million scientists between 1900 and 2020, we examine how knowledge mobility relates to innovation, learning, collaboration, and productivity. We find that scientists who continue to explore across fields sustain innovative contributions throughout their careers, whereas those who remain specialized exhibit the documented decline in innovation with age. Generalists are less anchored to the literature of their training, more likely to work independently, and more likely to collaborate with other generalists. Teams with a greater share of generalists produce more innovative research, even after accounting for knowledge diversity. Yet generalists publish fewer papers and are becoming increasingly rare.

Together, these findings suggest that innovation depends not only on the accumulation of knowledge through training and experience but also on the capacity for intellectual renewal. Yet the scientists most able to sustain innovation throughout their careers publish fewer papers and are becoming increasingly rare, revealing a tension between the longevity of scientific careers and the longevity of scientific innovation.

**Measuring career-long field exploration**

Scientists differ substantially in how they explore fields. Existing measures of interdisciplinarity typically count the number of fields represented in a scientist's publications or quantify their diversity using entropy-based indices (21–23). While useful for individual papers or short periods of time, these measures increase mechanically with career length and publication volume, making comparisons across scientists difficult. To overcome this limitation, we draw on scaling approaches widely used in complex systems research (24, 25). Just as Heaps' law characterizes how vocabulary grows with text length (26), we characterize how the cumulative number of fields explored $F$ grows with the number of publications $N$. We find that field exploration follows a power-law relationship, $F \sim N^{\beta}$. The scaling exponent ($\beta$) provides a scale-independent measure of how rapidly scientists expand into new fields as their careers progress. Across all analyzed scientists, the power-law model provides a good description of career trajectories (mean $R^2 \approx 0.8$). Moreover, scientists' relative rankings by ($\beta$) remain highly robust across alternative field classification resolutions (see SI for the detailed analysis).

Figure 1 illustrates this framework using two representative scientists. Herbert Simon, whose work spanned cognitive psychology, economics, computer science, and philosophy, exhibits a high scaling exponent ($\beta = 0.72$), indicating continual expansion into new fields throughout his career. In contrast, the physicist Giovanni Abbiendi exhibits a substantially lower exponent ($\beta =$ 0.23), reflecting concentration within a narrow set of fields despite a long publication record. Consistent with these differences, Simon's total publications are distributed relatively evenly across many fields, whereas Abbiendi's work is concentrated in a small number of fields (Fig. 1B). We therefore interpret ($\beta$) as a measure of career-long field exploration, with high-$\beta$ scientists representing generalists and low-$\beta$ scientists representing specialists.

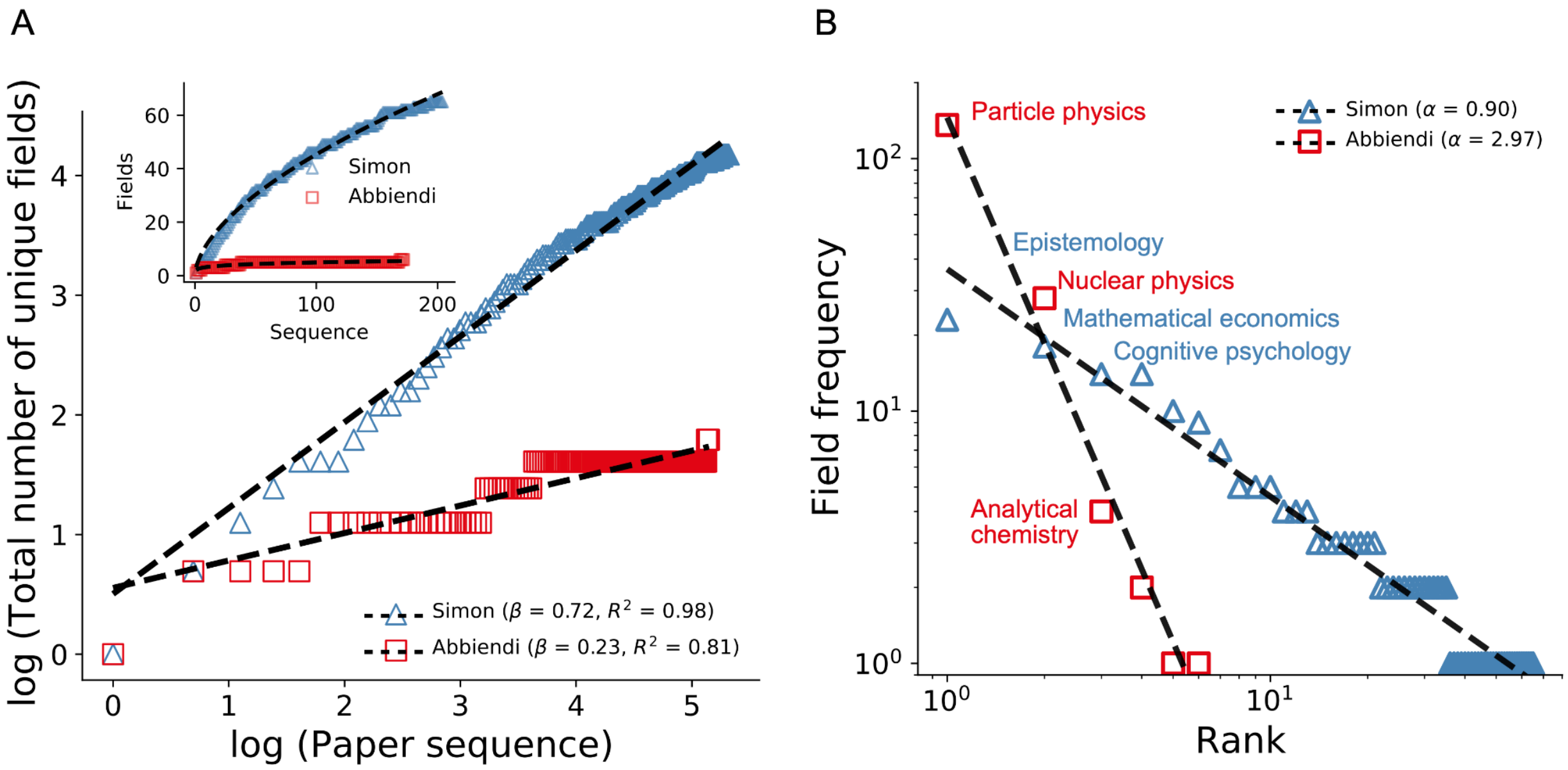

**Figure 1. Measuring career-long field exploration.** (A) Cumulative number of unique fields explored as a function of publication sequence for two representative scientists: Herbert Simon (blue triangles), a generalist, and Giovanni Abbiendi (red squares), a specialist physicist. Each point represents one first-authored paper; both axes are log-transformed. Dashed lines show power-law fits ($F \sim N^{\beta}$), where $\beta$ quantifies the rate at which scientists expand into new fields as they publish more papers. Simon expands across fields rapidly ($\beta$ = 0.72, $R^2$ = 0.98), whereas Abbiendi remains concentrated within a narrower set of fields ($\beta$ = 0.23, $R^2$ = 0.81). Insets show the same data in linear space. (B) Rank-frequency distributions of fields for each scientist on log-log axes. Fields are ranked by publication frequency; dashed lines show Zipf-law fits with exponent $\alpha$. Larger $\alpha$ indicates greater concentration of effort in a small number of fields (Abbiendi, $\alpha$ = 2.97), whereas smaller α indicates a more even distribution of effort across fields (Simon, $\alpha$ = 0.90). Field labels are shown for the most frequent fields for each scientist.

## Generalists sustain innovation but publish less

We next examine whether differences in field exploration are associated with differences in innovation trajectories. Scientists were grouped into quartiles according to their field-exploration exponent $\beta$. For each career year, we calculated the probability that a publication belonged to the top 5% most disruptive papers as measured by the D index. Consistent with recent studies, disruptive innovation declines with career age for all groups (Fig. 2A). However, the rate of decline differs substantially across scientists. Generalists (top $\beta$ quartile) maintain consistently higher probabilities of producing disruptive work than specialists (bottom β quartile) throughout their careers. Notably, late-career generalists exhibit disruptiveness levels comparable to those of early-career specialists. These results suggest that innovation decline is not a universal feature of scientific careers but varies systematically with patterns of field exploration.

At the same time, generalists publish substantially fewer papers per year than specialists throughout their careers (Fig. 2B). Whereas specialists rapidly increase their publication rates during early career stages before reaching a plateau, generalists maintain lower and relatively stable publication rates over time. Together, these findings reveal a tradeoff between productivity and sustained innovation across scientific careers.

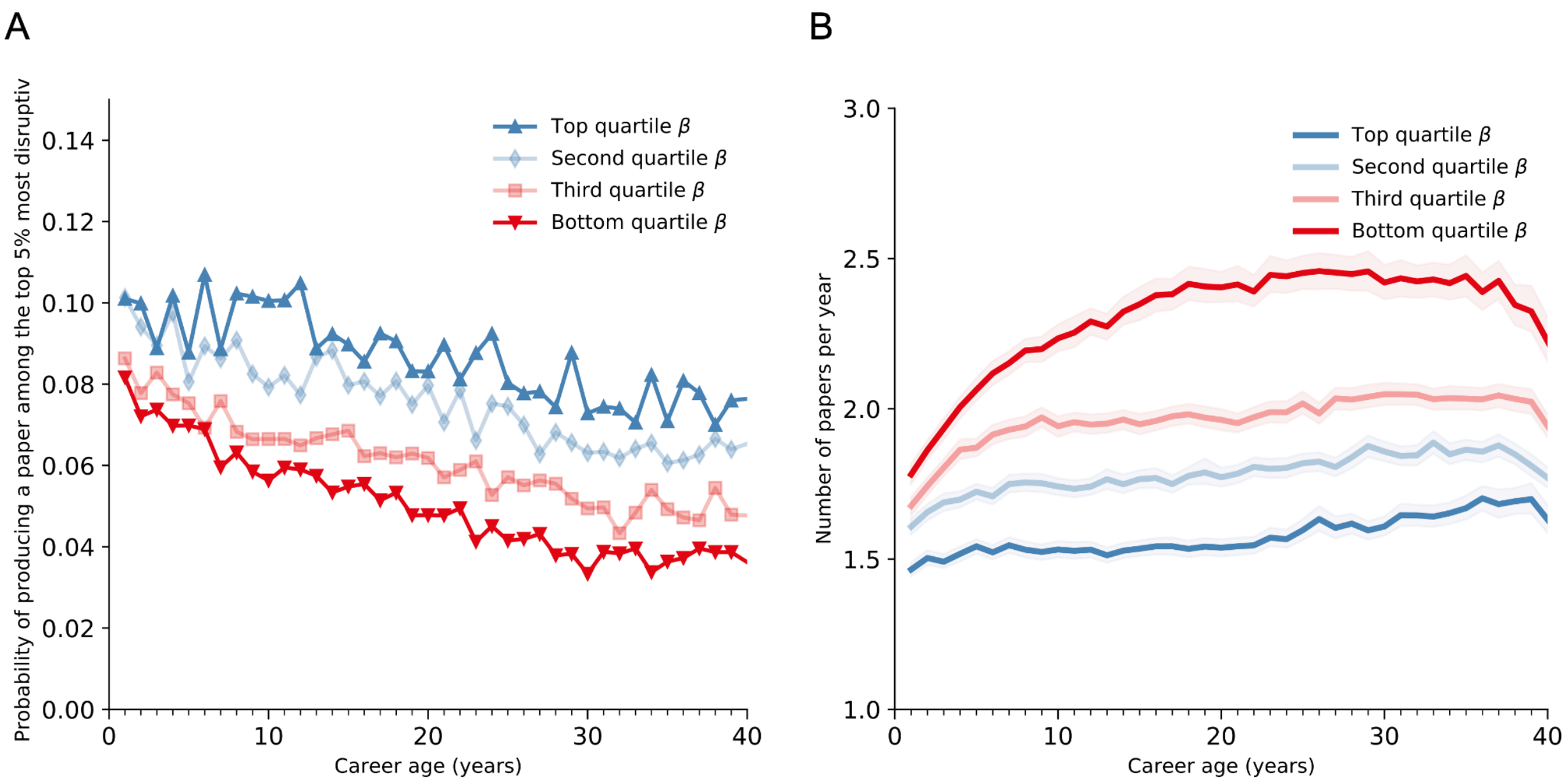

**Fig. 2. Generalists sustain innovation but publish less.** (A) Scientists are grouped into quartiles by their field-exploration scaling exponent $\beta$. For 87,419 scientists with careers spanning at least 40 years and five or more publications, we plot the probability that a publication belongs to the top 5% most disruptive papers (*D*-index, 10-year citation window) as a function of career age. Scientists in the highest $\beta$ quartile maintain higher disruptiveness throughout their careers than those in the lowest $\beta$ quartile. Notably, late-career generalists exhibit disruptiveness probabilities comparable to those of early-career specialists. (B) For 87,419 scientists with careers spanning at least 40 years and five or more publications, we plot mean annual publication output as a function of career age across quartiles of the field-exploration scaling exponent $\beta$. Specialists (bottom $\beta$ quartile) exhibit substantially higher publication rates than generalists (top $\beta$ quartile) throughout their careers. Shaded bands indicate 95% bootstrap confidence intervals.

## The collaboration and learning strategies of generalists

We identify each scientist's most frequently cited reference and assign it a relative career age measured from the scientist's first publication. In general, scientists remain anchored to references encountered earlier in their careers, likely reflecting the lasting influence of training and early intellectual experiences (3). However, generalists are less likely than specialists to remain anchored to these early references. Instead, they are more likely to use new references (Fig. 3B inset). These findings suggest that generalists continually update the intellectual foundations of their work rather than relying primarily on knowledge acquired early in their careers.

Generalists also exhibit distinct collaboration patterns. Specialists consistently produce a higher share of co-authored papers than generalists throughout their careers, suggesting that generalists maintain greater intellectual independence (Fig. 3A). At the same time, when generalists do collaborate, they tend to work with others who exhibit similarly high levels of field exploration (Fig. 3A inset). In other words, generalists preferentially collaborate with generalists, whereas specialists preferentially collaborate with specialists. These collaboration patterns appear beneficial: teams with a greater share of generalists are consistently more likely to produce highly disruptive work, even after accounting for the breadth of fields represented in their references (Fig. 3B; see SI for additional analyses).

One conventional explanation for the innovative advantage of generalists is that they combine knowledge from a larger number of fields. However, our findings suggest a different mechanism. The probability of producing a highly disruptive paper declines as the number of reference fields increases, indicating that drawing on more fields does not necessarily facilitate innovation and may instead create integration challenges. Moreover, even after accounting for the total number of reference fields represented in a paper, teams with a greater share of generalists remain consistently more disruptive than those with fewer generalists. These findings suggest that the innovative advantage of generalists arises not simply from their multidisciplinary experience, but from distinct learning and collaboration strategies that facilitate intellectual renewal.

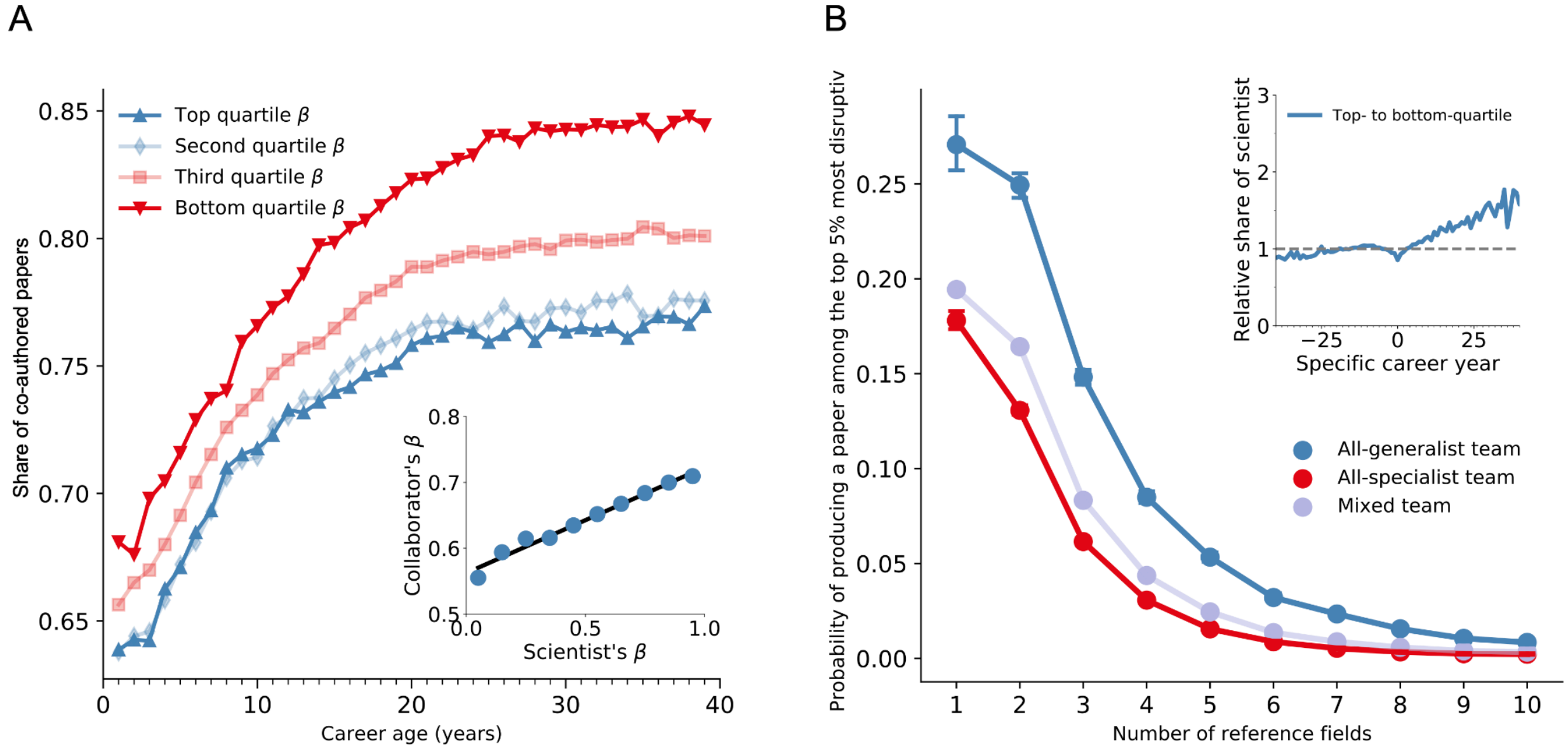

**Fig. 3. Generalists work more independently, collaborate with similar scientists, have less attachment, and contribute to more disruptive teams.** (A) For 87,419 scientists with careers spanning at least 40 years and five or more publications, we plot the share of single-authored papers as a function of career age across quartiles of the field-exploration scaling exponent $\beta$. Specialists (bottom $\beta$ quartile) consistently produce a higher share of collaborative papers than generalists (bottom $\beta$ quartile) throughout their careers. (Inset) For 2,688,209 scientists with five or more publications, we compare each scientist's $\beta$ with the mean $\beta$ of their collaborators. Scientists tend to collaborate with others of similar $\beta$, indicating assortative matching between generalists and specialists. (B) Scientists are classified as generalists ($\beta$ above the population median) or specialists ($\beta$ below the population median) and grouped into all-generalist, all-specialist, or mixed teams. We plot the probability that a paper belongs to the top 5% most disruptive papers as a function of the number of reference fields cited. Although disruptiveness declines with reference-field breadth, all-generalist teams consistently exhibit higher disruptiveness than mixed and

all-specialist teams across the full range examined. Error bars indicate 95% bootstrap confidence intervals. (Inset) For 2,688,209 scientists with five or more publications, we identify each scientist's most frequently cited reference and assign it a relative career age measured from the scientist's first publication. The curve shows the ratio of the share of top- to bottom-quartile $\beta$ scientists whose intellectual anchor originated at each relative career year. Values above 1 indicate that generalists are more likely than specialists to anchor their work to references from that career stage. Generalists are disproportionately anchored to more recent work, whereas specialists are disproportionately anchored to work encountered earlier in their careers.

## The reproduction of research styles and the decline of generalists

To understand how scientists develop their research styles, we compare each scientist's field-exploration exponent $\beta$ with that of their primary advisor. Scientists tend to resemble their advisors in $\beta$. This resemblance weakens for later supervisees and as the career age gap between supervisee and supervisor widens, suggesting that the influence of mentorship diminishes as supervisors advance in their careers (Fig. 4A). In other words, generalists are more likely to train generalists, and specialists are more likely to train specialists. This relationship suggests that field-exploration strategies are not randomly distributed across the scientific workforce as individual traits, but are instead socially reproduced through academic mentorship. As a result, differences in research style may persist and accumulate across scientific generations.

We find evidence consistent with this possibility. Across all ten disciplines examined, the share of top-quartile generalists declines substantially over the twentieth century (Fig. 4B). Scientists are assigned to disciplines based on the field of their first publication, allowing us to track long-term changes within disciplinary cohorts defined by training. The decline is particularly pronounced in the natural sciences, whereas the social sciences exhibit a more gradual decrease. These patterns are consistent with the widely documented rise of specialization and teamwork in modern science, especially within the natural sciences.

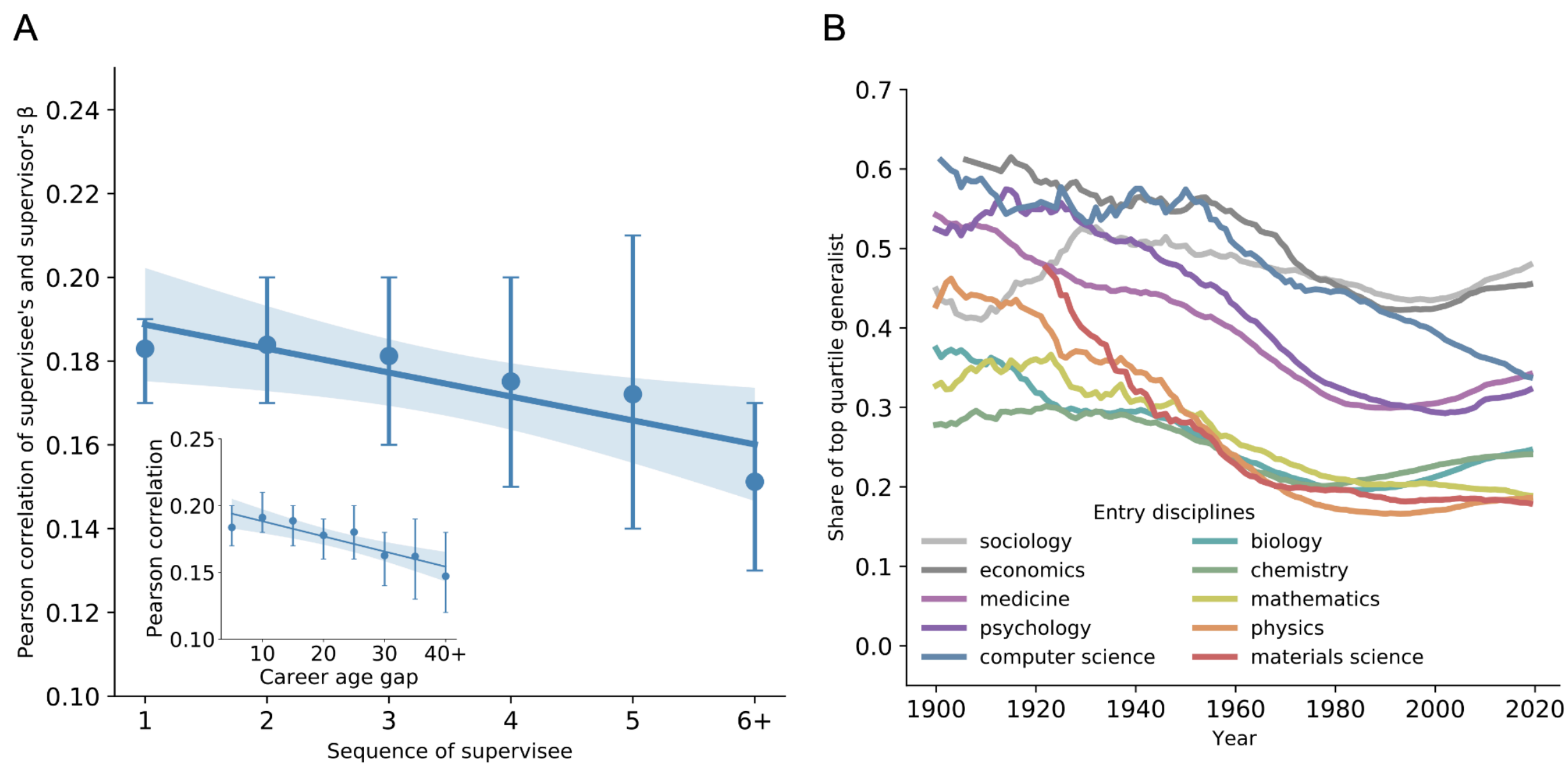

**Fig. 4. The reproduction of research styles and decline of generalists.** (A) For 87,840 scientists with five or more publications and with advisor information from Academic Tree, we compare each scientist's field-exploration exponent β with that of their primary supervisor. Scientists tend to resemble their supervisors in β, suggesting that research styles are transmitted across academic generations through mentorship. Transmission weakens for later supervisees (OLS slope = −0.006, $p < 0.05$) and as the career age gap between supervisee and supervisor widens (inset; OLS slope = −0.001, $p < 0.01$). Error bars indicate 95% confidence intervals (Fisher z-transformation). The shaded bands indicate the 95% confidence interval of the OLS fits. (B) For scientists grouped by entry discipline, we plot the share of top-quartile β scientists (generalists) among those entering scientific careers in each calendar year from 1900 to 2020, smoothed with a 20-year moving window. Across all ten disciplines shown, the prevalence of generalists declines over time, with particularly pronounced declines in the natural sciences.

## Discussion

Our findings challenge a common assumption underlying much of the literature on innovation and interdisciplinarity. Existing explanations often attribute the innovative advantage of generalists to their broader experience across fields (27). Yet we find that the breadth of knowledge alone cannot fully explain their performance. The probability of producing highly disruptive work declines as the number of reference fields increases, and teams with more generalists remain more innovative even after accounting for the breadth of knowledge represented in their work. These findings suggest that the advantage of generalists arises not simply from broad experience across fields, but from their ability to continually renew the intellectual foundations of their work.

We also find that the scientists most able to sustain innovation throughout their careers are not well represented in the scientific literature: they are increasingly rare, and publish fewer papers. At the same time, research styles are socially reproduced through mentorship, while modern science increasingly favors specialization and large-scale collaboration (5, 28). These findings reveal a tension between the longevity of scientific careers and the longevity of scientific innovation. Generalists sustain innovation by repeatedly exploring unfamiliar domains and maintaining greater intellectual independence, but these same behaviors are associated with lower productivity and weaker alignment with prevailing career incentives. Because research styles are reproduced through mentorship, these differences may persist across generations.

These findings have implications for science policy and research evaluation. Scientific training, hiring, promotion, and funding systems largely reward specialization. Even when interdisciplinarity is encouraged, it is often pursued through the assembly of large teams of specialists. Our findings suggest a different possibility: the innovative benefits of interdisciplinarity may arise not only from combining expertise across individuals, but also from intellectual mobility within individuals. Supporting intellectual mobility, independent research, and unconventional career trajectories may therefore be important not only for fostering interdisciplinarity, but also for sustaining innovation across increasingly long scientific careers.